# Stepped-height ridge waveguide MQW polarization mode converter monolithically integrated with sidewall grating DFB laser

Xiao Sun[1*], Weiqing Cheng[1], Song Liang[2], Shengwei Ye[1], Yongguang Huang[2], Ruikang Zhang[2], Bocang Qiu[3], Jichuan Xiong[4], Xuefeng Liu[4], John H. Marsh[1], Lianping Hou[1]

[1] *James Watt School of Engineering, University of Glasgow, Glasgow G12 8QQ, U.K.*
[2]*College of Physics, Jilin University, Changchun 130012, China*
[2] *Institute of Semiconductors, Chinese Academy of Sciences, No. A35, East Qinghua Road, Haidian District, Beijing 100083, China*
[3] *Institute of Atomic and Molecular Science, Shaanxi University of Science and Technology, Xian 712081, China*
[4] *School of Electronic and Optical Engineering, Nanjing University of Science and Technology, Nanjing 210094, China*
*Corresponding author: x.sun.2@research.gla.ac.uk

We report the first demonstration of a 1555 nm stepped-height ridge waveguide polarization mode converter monolithically integrated with a side wall grating distributed-feedback (DFB) laser using the identical epitaxial layer scheme. The device shows stable single longitudinal mode (SLM) operation with the output light converted from TE to TM polarization with an efficiency of >94% over a wide range of DFB injection currents ($I_{DFB}$) from 140 mA to 190 mA. The highest TM mode purity of 98.2% was obtained at $I_{DFB}$=180 mA. A particular advantage of this device is that only a single step of metalorganic vapor-phase epitaxy and two steps of III-V material dry etching are required for the whole integrated device fabrication, significantly reducing complexity and cost.

## 1. Introduction

Polarization mode controllers (PMCs) are of increasing importance in numerous applications in optical communication systems to manipulate the TE-TM polarization state of light [1]. Frequency conversation in periodically poled lithium niobate (PPLN) waveguides requires TM polarized light while most semiconductor edge emitting lasers operate in TE polarization. There has been a growing interest in integrating PMCs with devices using multiple-quantum-well (MQW) structures as the active region, such as polarization-dependent phase shifters (PD-PSs) [2] and laser diodes (LDs) [3, 4]. Previously reported PMCs based on MQW structures were able to convert pure TE- or TM-polarized light into an arbitrarily chosen state of polarization (SOP) but only with a 50% TE to TM polarization conversion efficiency (PCE) [2]. The PMC integrated with a 1550 nm Fabry-Perot (FP) LD reported in [3] realized a TE to TM PCE of only 80%. The crucial issue when integrating PMCs with MQW devices is the inherent birefringence of the MQW, which disturbs the optimal rotation of the SOP. The main mechanisms of SOP conversion in waveguides include the mode-coupling method [5], which exploits beating between two eigenmodes to enable polarization rotation along the PMC waveguide, and the mode-evolution method [6], which utilizes a change of the propagating mode inside the waveguide. The mode coupling approach to PMC design enables polarization conversion within a much shorter waveguide than the mode-evolution method. The mode coupling approach has therefore been proposed for integrating PMCs with MQW-based components such as LDs to reduce the internal loss caused by the strong exciton absorption inside the quantum well at the photoluminescence (PL) wavelength [7].

Several different PMC structures have been proposed such as waveguides using the reactive ion-etch (RIE) lag phenomenon [8], single-trench waveguides [9], angled-facet waveguides [10], and two-step waveguides [11]. These PMC devices use bulk material as the core layer in the waveguide and combine a high PCE with a short waveguide length. Nevertheless, relatively complicated butt-joint photonic integrated circuit (PIC) techniques involving re-growth are usually used to integrate PMCs with MQW-based devices. To simplify the monolithic fabrication, we have proposed an InP-AlGaInAs MQW-based sidewall grating (SWG) distributed feedback (DFB) laser with a stepped height waveguide PMC [12]. An optimized epitaxial design was proposed for integrating the DFB laser with the PMC waveguide. A series of full-wave simulations have been made to optimize the geometric parameters of PMC to obtain a high TE-TM PCE.

In this work, based on our simulation work in [12], an AlGaInAs MQW SWG DFB laser was fabricated and monolithically integrated with a stepped waveguide PMC based on the identical epitaxial layer (IEL) integration scheme for the first time. This approach needs only a single step of metalorganic vapor-phase epitaxy (MOVPE) step and two steps of III-V material etching. Compared with the conventional buried grating DFB lasers, the SWG DFB laser avoids the complicated etch and regrowth processes required to complete the laser structure after the grating definition. Compared with the butt-joint and selective area growth (SAG) PIC technologies, the IEL integration scheme eliminates time-consuming etch and regrowth steps. The DFB-PMC device reported here operates in a stable single longitudinal mode (SLM) and has a high PCE (>94%) over a wide range of DFB injection currents. The highest PCE obtained was 98%, which is consistent with the simulation results.

## 2. Device design and fabrication

The wafer structure used for the DFB-PMC is the same as that described in [12]. This wafer contains five 6-nm-thick compressively strained (+1.2%) AlGaInAs quantum wells (QWs) and six 10 nm-thick tensile strained (−0.3%) AlGaInAs quantum barriers (QBs). The room temperature PL peak of the QWs was located at a wavelength of 1530 nm. A 300-nm-thick 1.25Q InGaAsP (1.25Q means the PL wavelength of this material is 1.25 µm) layer is embedded below the MQW layer to increase the difference between the propagation constants of the two fundamental modes of the PMC so reducing the half-beat length ($L_\pi$) and increasing the PCE. The schematic of the DFB-PMC device is depicted in Fig. 1(a). It comprises a 1200 µm long shallow etched SWG DFB laser, a 50 µm long deep etched taper, and a 490 µm long PMC waveguide. The simulated reflection between the shallow etched DFB and deep etched taper sections is about $7\times10^{-6}$, which will have a negligible impact on DFB performance. Simulation also shows the excess optical loss of the taper is 1%, i.e., 0.044 dB, which includes the scattering and mode mismatch losses, and can also be neglected. The ridge waveguide of the DFB is 2.5 µm wide and 1.92 µm high. The grating period is 238 nm with a grating recess depth of 0.6 µm giving a 1.55 µm Bragg wavelength. The grating coupling coefficient κ was measured to be approximately 15 cm$^{-1}$ using the equation in [13]. A quarter wavelength shift section was inserted at the center of the DFB laser cavity to ensure SLM oscillation. The stepped-height PMC consisted of a ridge profile where $W_0$ and $W$ are the widths of the ridge waveguide and dry-etch corner, and $D_0$ and $D$ are the deep and shallow dry etched depths respectively. $D_0$ = 3.3 µm and $D$ =1.92 µm were chosen for the PMC waveguide as presented in Fig. 1(b). $D$ is the same as the DFB laser ridge height and can be precisely controlled because the 60 nm thick cladding AlGaInAs waveguide layer on top of the MQW layers acts as a dry etch stop layer when using a $CH_4/H_2/O_2$ recipe in an inductively coupled plasma (ICP) dry etch tool. To optimize the PMC width, a Full-Wave simulation was made using an FDTD software package. Figures 1(c) and (d) present the fundamental modes inside the taper tip cross-section and PMC waveguide. The PMC eigenmode is optimized to rotate the electric/magnetic fields through 45°. After propagating a half-beat length $L_\pi = \pi/(\beta_1-\beta_2)$ (where $\beta_1$ and $\beta_2$ are the propagation constants of the TE and TM eigenmodes respectively), the polarization is rotated through 90°, and the output mode becomes TM-polarized. The calculated effective modal indexes ($N_{eff}$) of the fundamental TE and TM modes in the PMC waveguide are 3.21109 and 3.20951 respectively. Figure 2 shows a contour plot of the calculated PCE and $L_\pi$ as a function of $W_0$ and $W$. The final optimum widths of the PMC waveguide chosen here are $W_0$ =1.38 µm and $W$ = 0.4 µm, which provide a high PCE (97.3%) and short $L_\pi$ (490 µm). The DFB-PMC fabrication process is presented in Fig. 3. The wafer is grown on an InP substrate by MOVPE (Fig. 3(a)). The DFB grating and PMC first step shallow etched waveguide pattern is defined by e-beam lithography (EBL) using negative tone Hydrogen Silsesquioxane (HSQ) photoresist which acts as both an EBL resist and ICP dry etching hard mask in height (Fig. 3(e)). Finally, the HSQ resist was removed with hydrofluoric acid in Fig. 3(f). The fabrication workflow is depicted in Fig. 3(g); only a single step of MOVPE and two steps of dry etching are required for the whole integrated device. SEM images of the DFB grating, taper, and PMC waveguide are presented in Fig. 4(a)-(c). The subsequent deposition of SiO$_2$ and HSQ passivation layers, SiO$_2$ window opening, P-contact deposition, substrate thinning, and N-contact deposition are the same as for conventional LD fabrication and can be referred to [14]. An optical microscope picture of the completed DFB-PMC device is depicted in Fig. 4(d). Finally, the devices were mounted epilayer up on a copper heat sink on a Peltier cooler. The heat sink temperature was set at 20°C and the devices were tested under CW conditions.

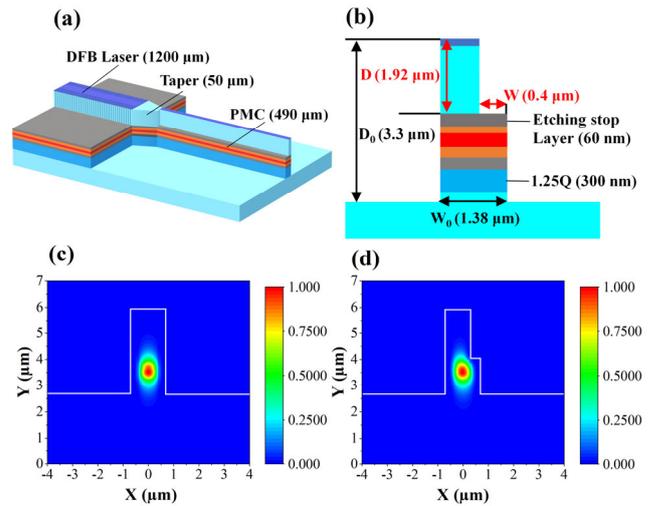

Fig. 1. (a) Schematic of the monolithic DFB-PMC device, (b) cross-section structure of the PMC; (c)-(d) the fundamental eigenmodes in the taper tip cross-section (c), and in the PMC stepped-height ridge waveguide (d).

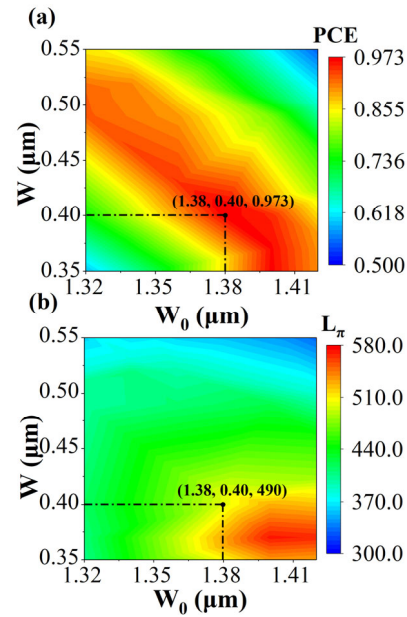

Fig. 2. (a) -(b) Calculated maximum PCE (a), and corresponding $L_\pi$ (b) as a function of waveguide width ($W_0$) and corner width ($W$).

Fig. 3(b). In Fig. 3(c), the ridge is first etched to a depth of 1.89 µm in an ICP system using a $Cl_2/CH_4/H_2/Ar$ gas mixture, where the etching rate for InP, InGaAsP, and AlGaInAs is about 182 nm/minute. Then, the gas recipe was changed to $CH_4/H_2/O_2$ and the ridge waveguide was continuously etched to 1.92 µm height, with an etch rate for InP and InGaAsP of about 78 nm/minute, while that of the upper 60 nm AlGaInAs layer was 3 nm/minute which is a 26-fold selectivity with respect to InGaAsP and InP. After the shallow ridge waveguide was etched, both the DFB grating and PMC's first step waveguide were protected using HSQ, and the second PMC deep etched ridge waveguide was defined by EBL as shown in Fig. 3(d). A second ICP etch with $Cl_2/CH_4/H_2/Ar$ was used to etch the PMC ridge waveguide to 3.3 µm

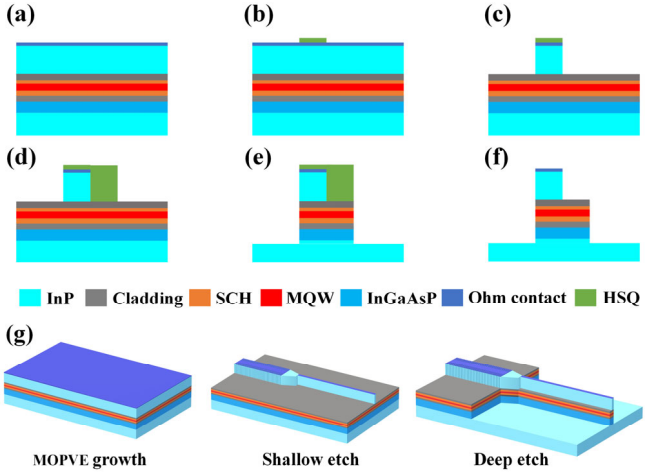

Fig. 3. Fabrication procedures: (a) MOVPE wafer, (b) EBL to write the laser and PMC first step waveguide, (c) ICP shallow etching, (d) EBL to write the second step waveguide of PMC, (e) ICP deep etching, (f) HSQ strip off, (g) workflow of the monolithic DFB-PMC device fabrication

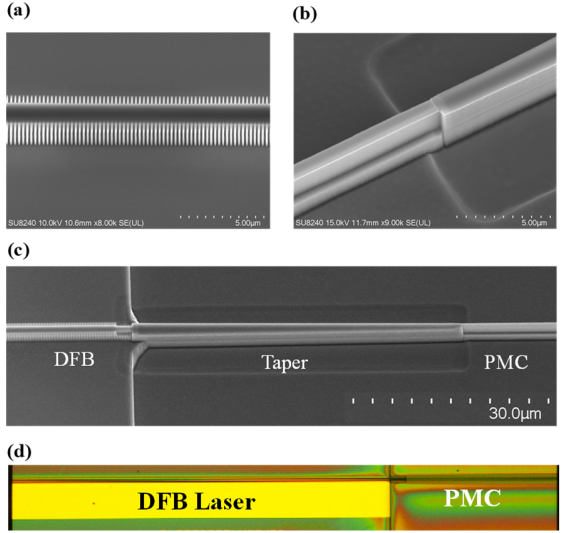

Fig. 4. (a) SEM image of DFB SWGs, (b) SEM image at the interface between PMC and taper, (c) SEM images of DFB-PMC device, (d) microscope picture of the DFB-PMC device

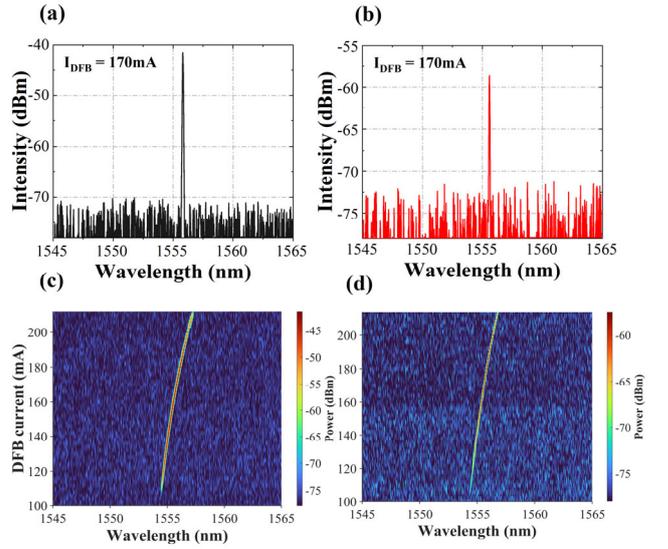

Fig. 5. (a)-(b) Measured optical spectrum from (a) DFB LD rear facet and (b) PMC section output facet at 170 mA. (c)-(d) 2D optical spectrum at (c) DFB LD rear facet and (d) PMC section output facet.

## 3. Device measurements

The measured internal loss for PMC waveguide at 1550 nm wavelength is 9/cm. Figure 5(a) shows the optical spectrum from the rear side of the DFB section at $I_{DFB}$=170 mA. The spectrum was measured with a resolution bandwidth (RBW) of 0.06 nm. The peak lasing wavelength is 1555.81 nm with a side mode suppression ratio (SMSR) of at least 29 dB, the measurement of which is limited by the amount of power coupled into the lensed fiber from the rear side of the DFB LD.

Figure 5(b) presents the optical spectrum from the PMC section output side. The attenuation of output power at the PMC facet is due to the strong exciton absorption inside the PMC waveguide when the propagating light wavelength is close to the PL wavelength of the MQW core (1530 nm). Figure 5(c) and (d) present 2D optical spectra measured from the DFB rear side and PMC output side respectively over a range of $I_{DFB}$ from 100 mA to 220 mA. There is stable SLM operation over a wide range of $I_{DFB}$ without any longitudinal mode hops. The threshold current, $I_{DFB}$, is 104 mA. The average current-induced wavelength redshift coefficient was found to be 0.020nm/mA. The setup of the SOP measurement is shown in Fig. 6(a). The DFB-PMC device was mounted on a thermoelectric cooler (TEC) and temperature controlled at 20°C as stated previously. The output light from PMC was coupled to a lensed polarization maintaining (PM) fiber and transmitted to a polarimeter to measure the SOP. Both the current driver and the polarimeter were controlled by a computer through the general-purpose interface bus (GPIB) interface. First, the SOP from the DFB laser rear facet was measured at $I_{DFB}$ from 104 mA to 210 mA, and the Stokes vector was constant at $(S_1, S_2, S_3)$ = (0.998, 0.05, 0.04), where $S_1, S_2, S_3$ are the S-parameters of the SOP. As expected, the light was TE-polarized with a purity of 99.8%. The SOP at the PMC section output facet is depicted in Fig. 6(b). The $S_1$ parameter was found to be <–0.94 within the range 140 mA < $I_{DFB}$ < 190 mA. Outside this this range the $S_1$ parameter still remained <–0.8. This is because the SMSR from the PMC output facet is lower than the SMSR sensitivity of the polarimeter when $I_{DFB}$ is <140 mA or >190 mA. The maximum $S_1$ is detected at $I_{DFB}$ = 180 mA where the SOP is (–0.982, 0.08, 0.17) representing a TM purity of 98.2%. This result is nearly the same as the simulated PCE of 97.3% Fig. 6(c) shows the SOP on the Poincaré sphere for the DFB laser rear facet and the PMC output facet with $I_{DFB}$ = (140 – 190) mA. The SOP at $I_{DFB}$ =

140 mA corresponds to the inside point of the "U" shape curve on the Poincaré sphere.

We note the fabricated PMC length should be kept as close as possible to the designed value by precise control of the cleaving. Here a LOOMIS LSD-100 cleaving tool was used with a cleaving accuracy of ±1 μm. The resulting variation in the PCE is less than 0.1%, confirming the tool meets the required tolerance. We also comment that, in order to increase the output power of the device, quantum well intermixing (QWI) could be used to blueshift the bandgap in the PMC section and reduce its absorption loss.

## 4. Conclusion

A novel monolithically integrated DFB-PMC device has been fabricated based on a SWG and IEL structure for the first time. The PMC is designed as a stepped height ridge waveguide. A major advantage of the design is that only a single MOVPE step and two dry-etch steps are required to fabricate the device. The device operates in a stable SLM with a current-induced wavelength redshift coefficient of 0.020 nm/mA. The PMC has a high TM purity (> 94%) over a wide $I_{DFB}$ range (140-190 mA), and a maximum TM mode purity of 98.2% was measured at $I_{DFB}$ = 180 mA.

**Funding.** This work was supported by the U.K. Engineering and Physical Sciences Research Council (EP/R042578/1) and the Chinese Ministry of Education collaborative project (B17023).

**Acknowledgments.** We would like to acknowledge the staff of the James Watt Nanofabrication Centre at the University of Glasgow for their help in fabricating the devices.

**Disclosures**. The authors declare no conflict of interest.

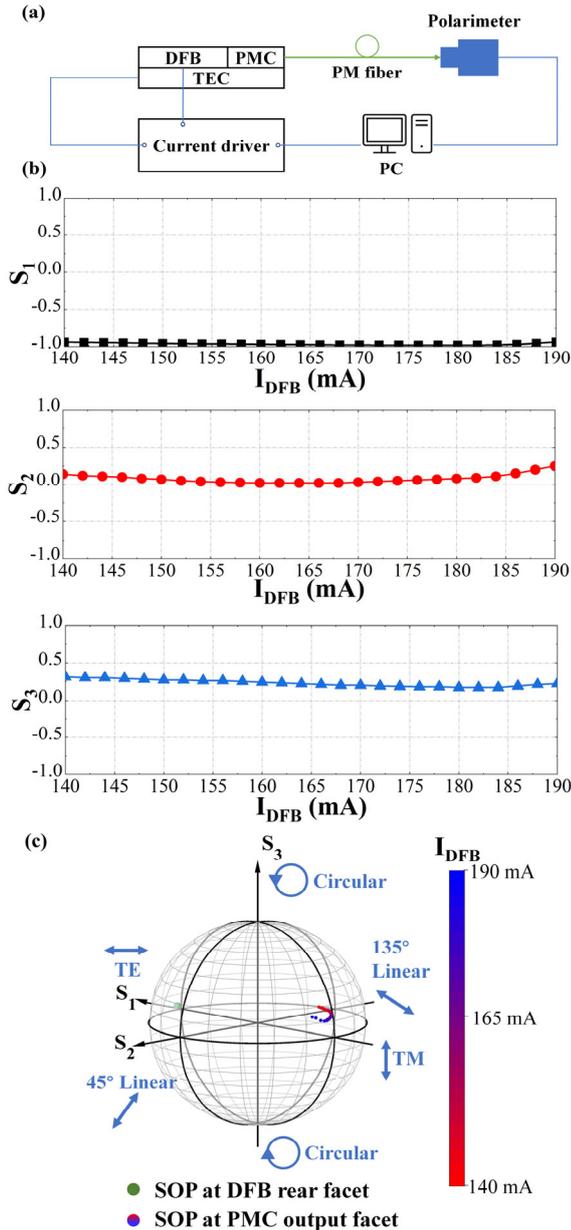

Fig. 6. (a) SOP measurement setup, (b) S-parameters of the SOP at PMC output facet side as a function of IDFB, (c) SOP on the Poincaré sphere with IDFB from 140 mA to 190 mA.